\documentclass[npg]{copernicus}

\def\dsize{\displaystyle}
\def\E{\mathop{\rm E}\nolimits}
\def\Ro{\mathop{\rm Ro}\nolimits}

\def\Re{\mathop{\rm Re}\nolimits}
\def\Ra{\mathop{\rm Ra}\nolimits}

\def\Pr{\mathop{\rm Pr}\nolimits}
\def\q{\mathop{\rm q}\nolimits}
\usepackage{bm}
\usepackage{epsfig}

\begin{document}

\title{Direct and inverse cascades in  geodynmo}

\author[1]{P. Hejda}
\author[2]{M. Reshetnyak}

\affil[1]{Geophysical Institute, Academy of Sciences, 141 31
Prague, Czech Republic}
\affil[2]{Institute of the Physics of the Earth, Russian Acad.~Sci,
 123995 Moscow, Russia}

%% The [] brackets identify the author to the corresponding affiliation, 1, 2, 3, etc. should be inserted.

\runningtitle{Direct and inverse cascades}

\runningauthor{Hejda \& Reshetnyak}

\correspondence{M.Reshetnyak\\ (m.reshetnyak@gmail.com)}

\received{}
\pubdiscuss{} %% only important for two-stage journals
\revised{}
\accepted{}
\published{}

%% These dates will be inserted by the Publication Production Office during the typesetting process.

\firstpage{1}

\maketitle

\begin{abstract}
The rapid rotation of the planets causes cyclonic thermal
turbulence in the cores, which can be responsible for the generation of the large-scale
 magnetic fields observed out of the planets. We consider the model which lets us to reproduce
 the typical features of the small-scale geostrophic flows in the
 physical and wave spaces. We present estimates of kinetic and magnetic  energies
 fluxes as a function of the wave number. The joined
 existence of the forward and inverse cascades are demonstrated.
 We also consider mechanism of the magnetic field saturation at the end of the kinematic dynamo regime.
\end{abstract}

\introduction
%% \introduction[modified heading if necessary]

Many astrophysical objects such as galaxies, stars, the earth, and some of the planets  have
large scale magnetic fields that are believed to be generated by a
common universal mechanism - the conversion of kinetic energy into
magnetic energy in a turbulent rotating shell. The details, however,
 - and thus the nature of the resulting field - differ greatly.
  The challenge for the dynamo theory see, e.g., \citep{HolRud}, is to provide a
model that can explain the visible features of the field with
realistic assumptions on the model parameters. Calculations for the
entire  planet are done either with spectral models \citep{KonoRoberts}
or finite volume methods \citep{HR04,HaHa} and finite difference \citep{Kag}, and have demonstrated beyond reasonable
doubt that the turbulent 3D convection of the conductive fluid
 can generate a large scale magnetic field
similar to the one observed out of small random fluctuations.
However, both these methods cannot cover the
enormous span of scales required for a realistic parameter set.
Even for the geodynamo (which is quite a modest on astrophysical scales case) the hydrodynamic Reynolds
 number estimated on the west drift velocity is $\Re\sim 10^9$. In addition planets are the rapidly rotating bodies. So,
 the time scale of the large-scale convection in the Earth's core is
$\sim 10^3$ years, during which the planet itself makes $\sim 10^6$
revolutions (in other words the Rossby number $\Ro\sim 10^{-6}$).
As a result, there is an additional  spatial scale $\sim \E^{-1/3}L$, where  $\E\sim 10^{-15}$ is
 the Ekman number \citep{Ch, Busse70}, associated with the cyclonic structures elongated along axis of rotation,
which is quite larger then the Kolmogorov's dissipation
  scale $l_d\sim \Re^{-4/3}{ L}$ however which is  still too small to be resolved
 in the numerical simulations with present resolution $l\sim (10^{-3}\div 10^{-2}){ L}$ with $ L$ for the large scale.

  Presence of the rapid rotation leads not only to the change of the spatial uniform isotropic Kolmogorov like
     solution to the quasi-geostrophic (magnetostrophic) form  but to rather more fundamental consequences.
 Rapid rotation leads to degeneration of the third dimension (along the axis of rotation) and can cause
  inverse cascade in the system. The inverse cascades is a well-known phenomenon
    in the two-dimensional turbulence and is a good example of
   self-organization, when the large-scale structures are feeded  by the small-scales turbulence \citep{Kr, Tabel}.
    So far, that the quasi-geostrophic state  formally still is a three dimensional (the vector fields have three
     components) behavior of such systems can differ from the two-dimensional flows as well as from the three dimensional
      ones. It can happen, that quasi-geostrophic turbulence can exhibit simultaneously features
      similar to the both extreme cases: 2D and 3D. Further we consider behavior of the fluxes of the energy in the
       wave space for the well-known in geodynamo regimes based on the Boussinesque thermal convection.
         For simplicity we consider the Cartesian geometry which is
        simpler for modeling of the rapidly rotating dynamo systems and was used in many researches on geodynamo
\cite{Roberts99, JR, Buffett}.

\section{Dynamo equtions}
\subsection{Equations in physical space}
The geodynamo equations for the incompressible fluid
  ($\nabla\cdot{\bf V}=0$) in the volume of the scale $L$
 rotating with the angular velocity $\Omega$  in the Cartesian system of
  coordinates $(x,\,y,\,z)$
 in its traditional dimensionless form can
be written as follows:
\begin{equation}
\begin{array}{l}\dsize
   \frac {{\partial}{\bf B}}{ {\partial t}} ={ \nabla}\times \left({\bf V}\times{\bf B}\right)+
\q^{-1} \Delta {\bf B} \\
\dsize
    \E\Pr^{-1}\left[\frac {{\partial} {\bf V}} {\partial t}+ \left({\bf
V}\cdot \nabla\right) {\bf V}\right] = -\nabla { P}  -{\bf
{1}_z}\times{\bf V} + \\ \qquad\qquad \Ra { T} \,z{\bf{1}_z}+
\left(
\nabla\times {\bf B}
\right)\times {\bf B}+
 \E\Delta {\bf V}
\\ \dsize
\frac{\partial { T}} {\partial t}+\left({\bf V}\cdot\nabla\right)
\left({ T}+{ T}_0\right)= \Delta{T}.
\end{array}\label{sys0}
\end{equation}
The velocity $\bf V$, magnetic field $\bf B$, pressure
$P$ and typical
diffusion time $t$ are measured in units of
$ \kappa/L$, $\dsize \sqrt{2\Omega\kappa\mu\rho};$, $\rm
\rho\kappa^2/L^2$ and $\rm L^2/\kappa$ respectively, where $\kappa$ is
thermal  diffusivity, $\rho$ is density, $\mu$ permeability,
$\dsize \Pr=\frac{\kappa}{
\nu}$ is the Prandtl number,
  $ \dsize \E = \frac{\nu}{ 2\Omega L^2}$ is the
Ekman number, $\nu$ is kinematic viscosity,
$\eta$ is the magnetic diffusivity, and ${\q}=\kappa/\eta$ is the
Roberts number.
$\dsize \Ra
=\frac{\alpha g_0\delta T {L}}{  2\Omega\kappa}$ is the modified
Rayleigh number, $\alpha$ is the coefficient of volume expansion,
$\delta T$ is the unit of temperature, see for more details \cite{Jones},
 $g_0$ is the gravitational acceleration, and $T_0=1-z$ is the heating below.
 The problem is closed with the periodical boundary conditions in the $(x,\, y)$ plane. In $z$-direction we use
  simplified conditions  \citep{Cattaneo}:
$T_0=0$,
  $\dsize V_z={\partial V_x\over \partial z}=
 {\partial V_y\over \partial z}=0$,   $\dsize B_x=B_y={\partial B_z\over \partial z}=0$ at $z=0,\,1$.

\subsection{Equations in wave space}
To solve the problem (\ref{sys0})
 we apply pseudo-spectral approach
\citep{Or} frequently used in geodynamo simulations \citep{JR, Buffett}.
 Equations are solved in the wave space. To
calculate the non-linear terms one needs
 make the inverse Fourier transform, then calculate a product in the physical space,
   make a Fourier transform of the
  product, and finally calculate a derivatives in the wave space.
After elimination of the pressure using conditions of
 the free divergency ${\bf k}\cdot {\bf V}=0$, ${\bf k}\cdot {\bf B}=0$ we come to
\begin{equation}\begin{array}{l}\dsize
\left[ {{\partial }{\bf B} \over {\partial t}}+\q^{-1}k^2{\bf B}
\right]_{\bf k}=
 \left[{ \nabla}\times \left({\bf V}\times{\bf B}\right)\right]_{\bf k} \\ \\

\dsize
 \E\left[\Pr^{-1}
 {\partial {\bf V}\over\partial t}
  +k^2{\bf V}\right]_{\bf k}
= {\bf k} {\cal P}_{\bf k}+{\bf F}_{\bf k}
\\ \\ \dsize
\left[ {\partial { T}\over\partial t}+k^2 T\right]_{\bf k}=
-\left[\left({\bf V}\cdot\nabla\right){ T}+V_r\right]_{\bf k}
\end{array}\label{sys_s}
\end{equation}
with
\begin{equation}\begin{array}{l}\dsize

{\cal P}_{\bf k}=-{ {\bf k}\cdot {\bf F}_{\bf k}\over k^2}, \qquad
\dsize k^2=k_\beta k_\beta,\qquad \beta=1\dots 3
\\\\
\dsize {\bf F}_{\bf k}=\Big[ \Pr^{-1} {\bf V} \times
\left(\nabla\times {\bf V} \right)+ \Ra T{\bf{1}_z}-\\ \qquad {\bf
1_z}\times {\bf V}+ \left({\bf B }\cdot \nabla\right){\bf B}
 \Big]_{\bf k}.
\end{array}\label{P123}
\end{equation}
For integration in time we use explicit
 Adams-Bashforth (AB2) scheme for the non-linear terms.
 The linear terms are treated using the Crank-Nicolson (CN) scheme. To resolve
  the diffusion terms we use the known trick which helps to increase
   the time step significantly. Consider equation
\begin{equation}\begin{array}{l}\dsize
{\partial A  \over \partial t}+k^2A=U
\end{array}\label{tr1}
\end{equation}
rewrite it in the form
\begin{equation}\begin{array}{l}\dsize
{\partial A e^{k^0\gamma t} \over \partial t}=U\, e^{k^2\gamma t}
\end{array}\label{tr2}
\end{equation}
and then the CN scheme is applied.

The most time consuming part of our MPI code is FFT transforms. To make
our code more efficient we use various modifications of known FFT
transforms which take into account special kinds of the symmetry of
the fields. The optimal number of processors for the grids
$128^3$ is $n\sim 50$. The scalability tests
demonstrates even presence of the superacceleration for the number
of processors $<n$.

\begin{figure*}[t]
\vskip -4cm
\begin{minipage}[h!]{.45\linewidth}
\vspace*{2mm}
\begin{center}
\vskip -30cm
\includegraphics[width=40cm]{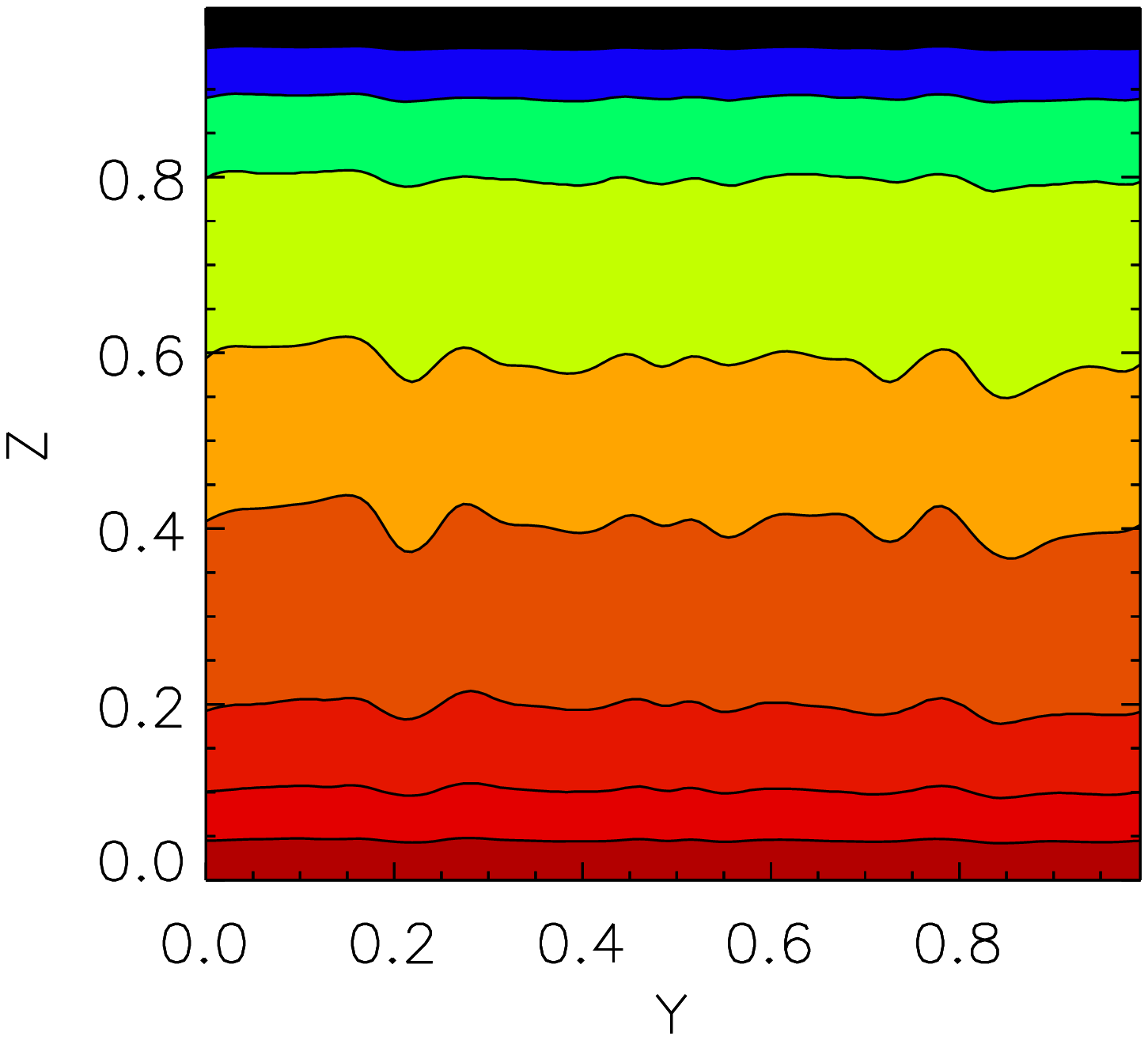}
\end{center}
\end{minipage}\hfill
\hskip -10cm
\begin{minipage}[h!]{.45\linewidth}
\vspace*{2mm}
\begin{center}
\vskip -30cm
\includegraphics[width=40cm]{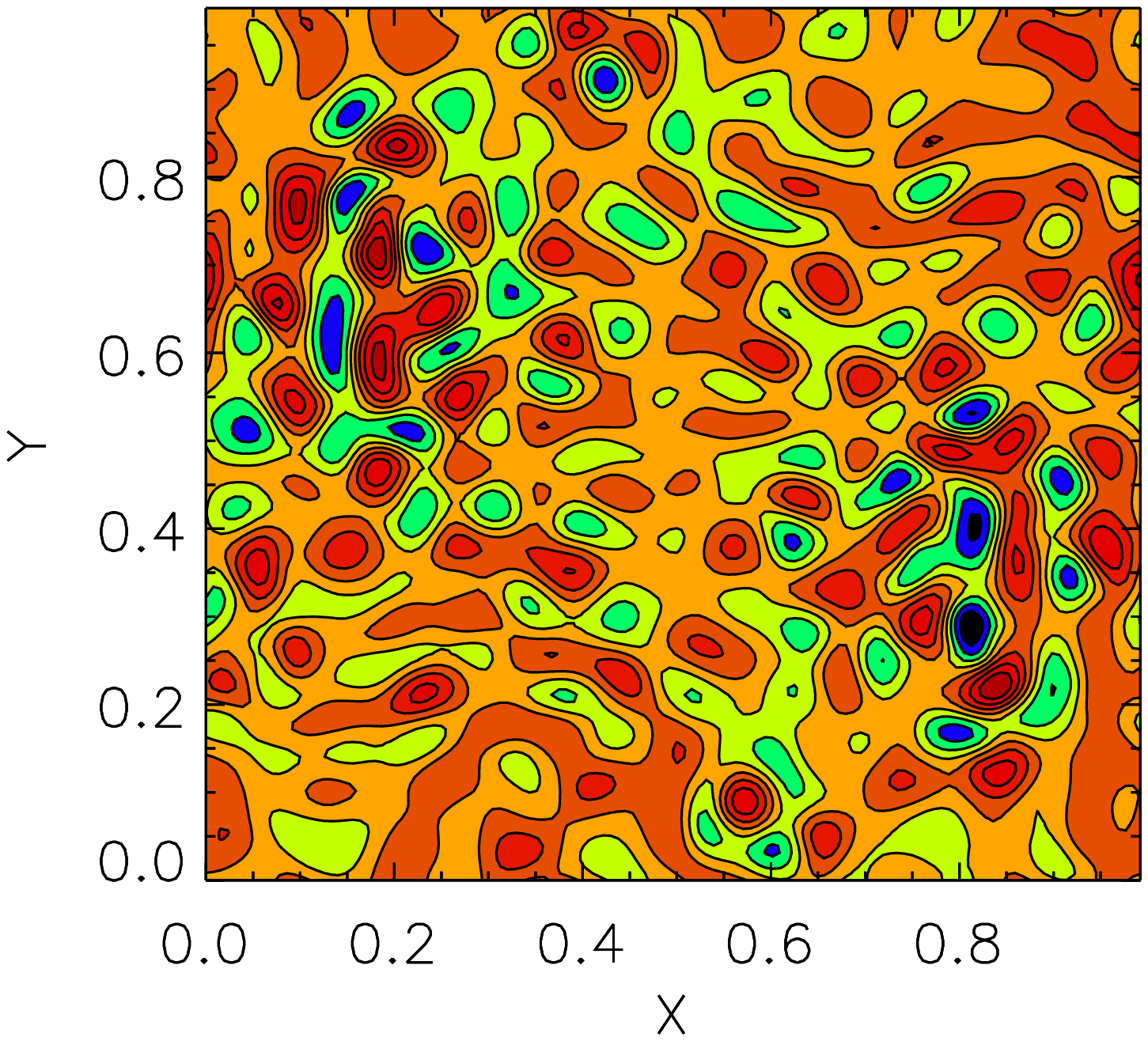}
\end{center}
\end{minipage}\hfill
\vskip 1cm
\caption{Regime R1. Sections of temperature distribution. . All sections correspond to
 the middle of the cube. The field ranges are  $(0,\, 1)$ -- left, $(0.46,\, 0.54)$ -- right.}
\label{Fig1}
\end{figure*}

\begin{figure*}[t]
\vskip -4cm
\begin{minipage}[h!]{.45\linewidth}
\vspace*{2mm}
\begin{center}
\vskip -30cm
\includegraphics[width=40cm]{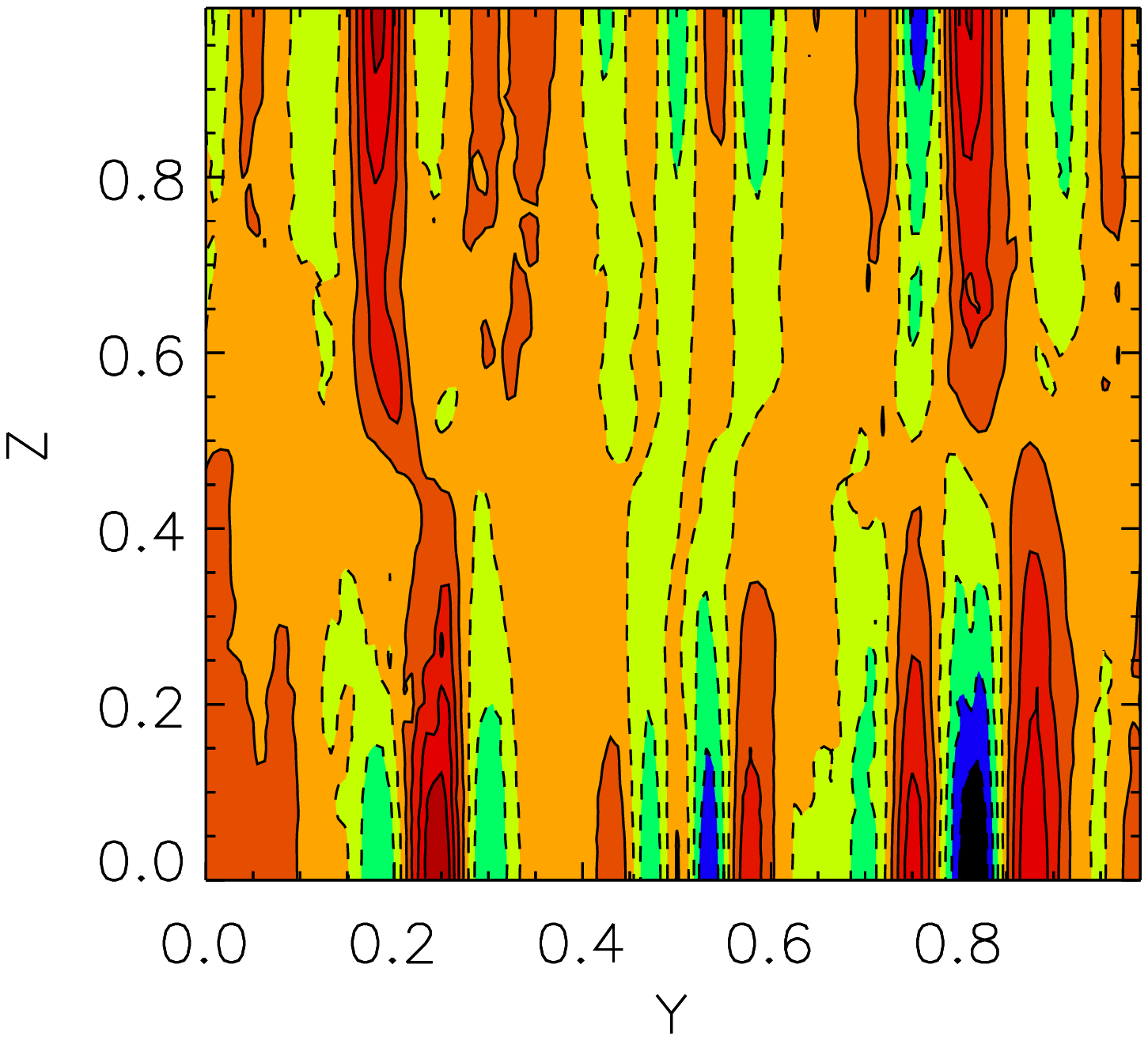}
\end{center}
\end{minipage}\hfill
\hskip -10cm
\begin{minipage}[h!]{.45\linewidth}
\vspace*{2mm}
\begin{center}
\vskip -30cm
\includegraphics[width=40cm]{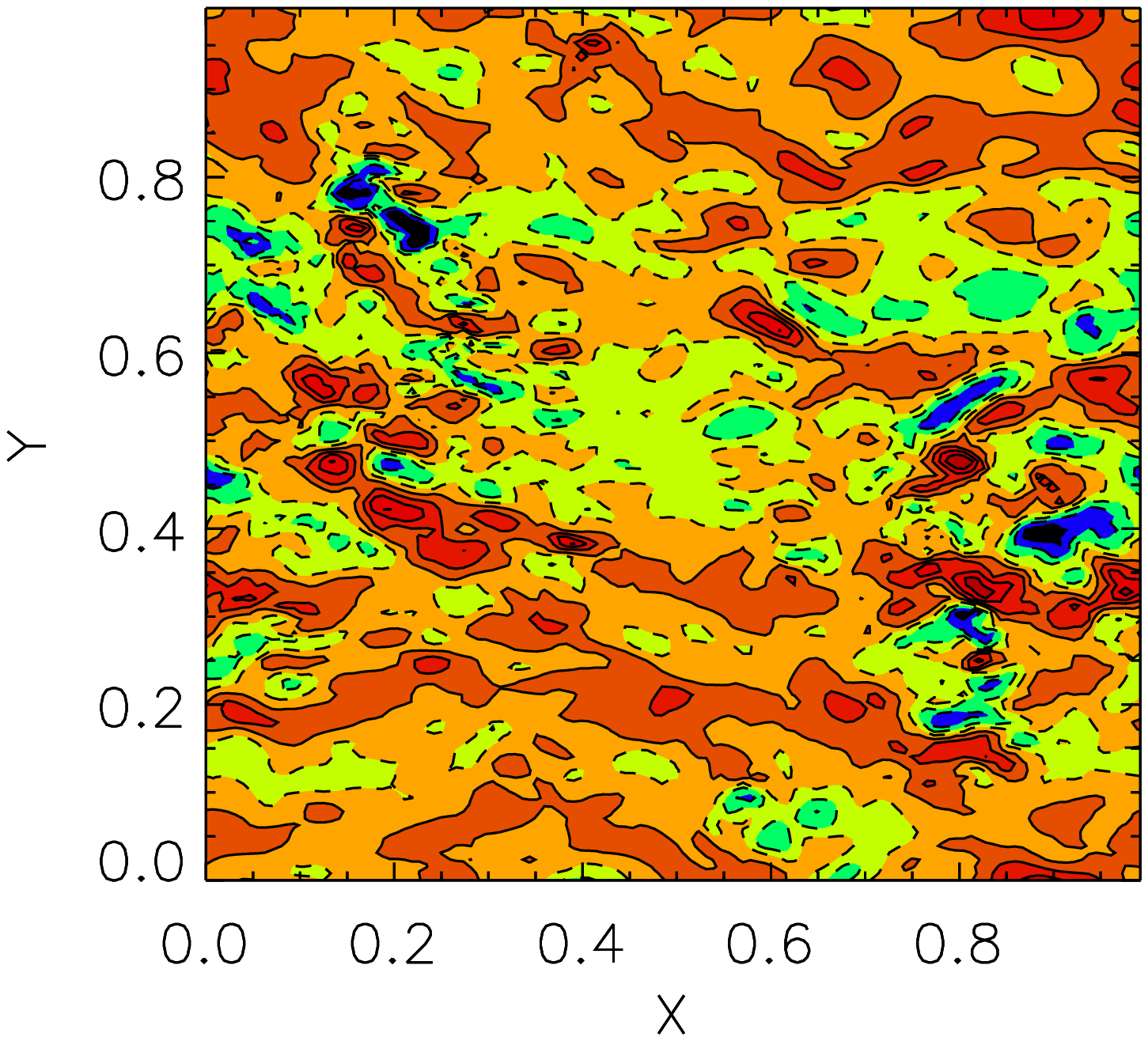}
\end{center}
\label{Fig2}
\end{minipage}\hfill
\vskip 1cm
\caption{Distribution of $V_x$-component of the velocity field with ranges $(-248,\, 253)$, $(-143,\, 144)$.}
\end{figure*}

\begin{figure*}[t]
\vskip -4cm
\begin{minipage}[h!]{.45\linewidth}
\vspace*{2mm}
\begin{center}
\vskip -30cm
\includegraphics[width=40cm]{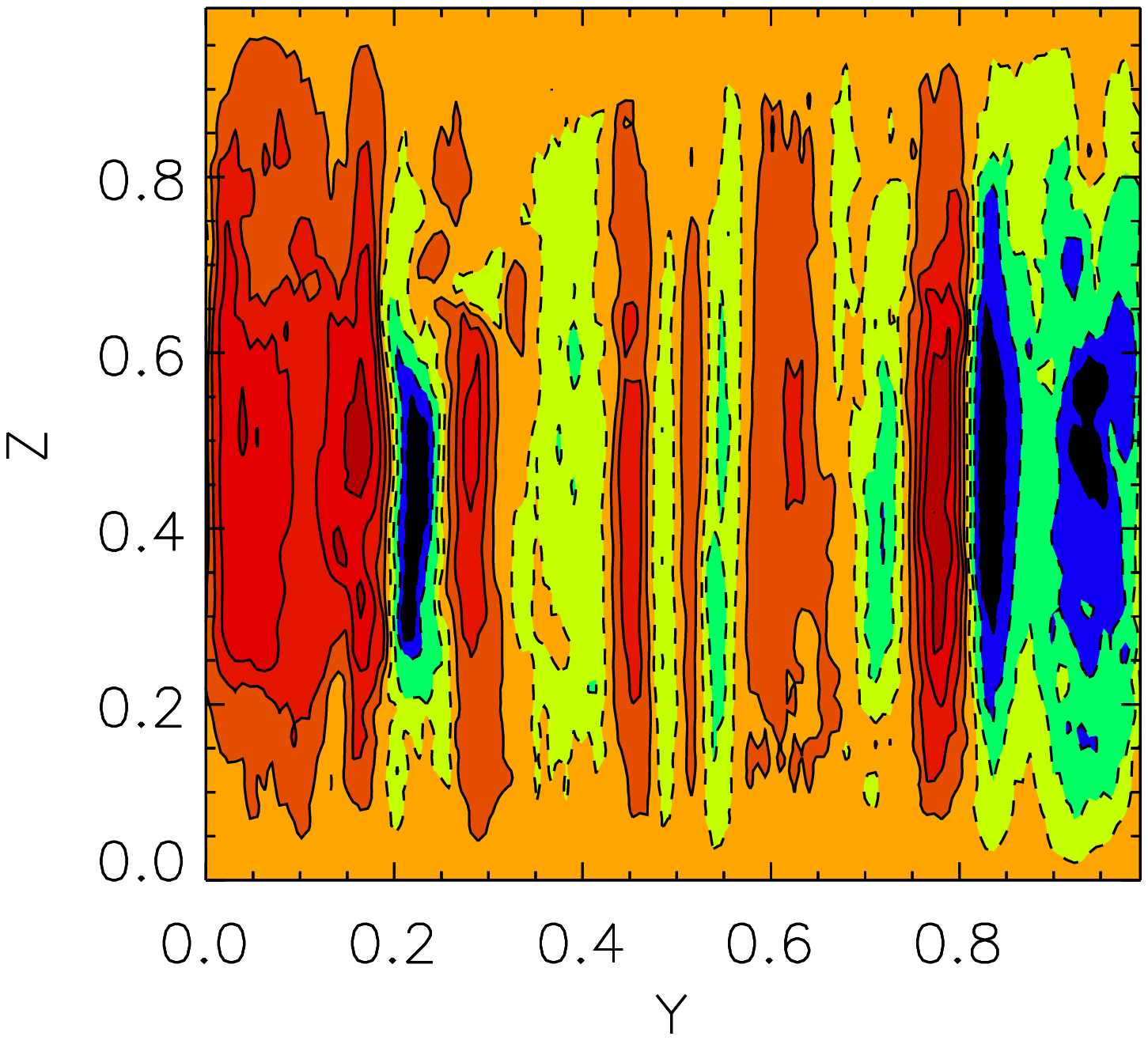}
\end{center}
\end{minipage}\hfill
\hskip -10cm
\begin{minipage}[h!]{.45\linewidth}
\vspace*{2mm}
\begin{center}
\vskip -30cm
\includegraphics[width=40cm]{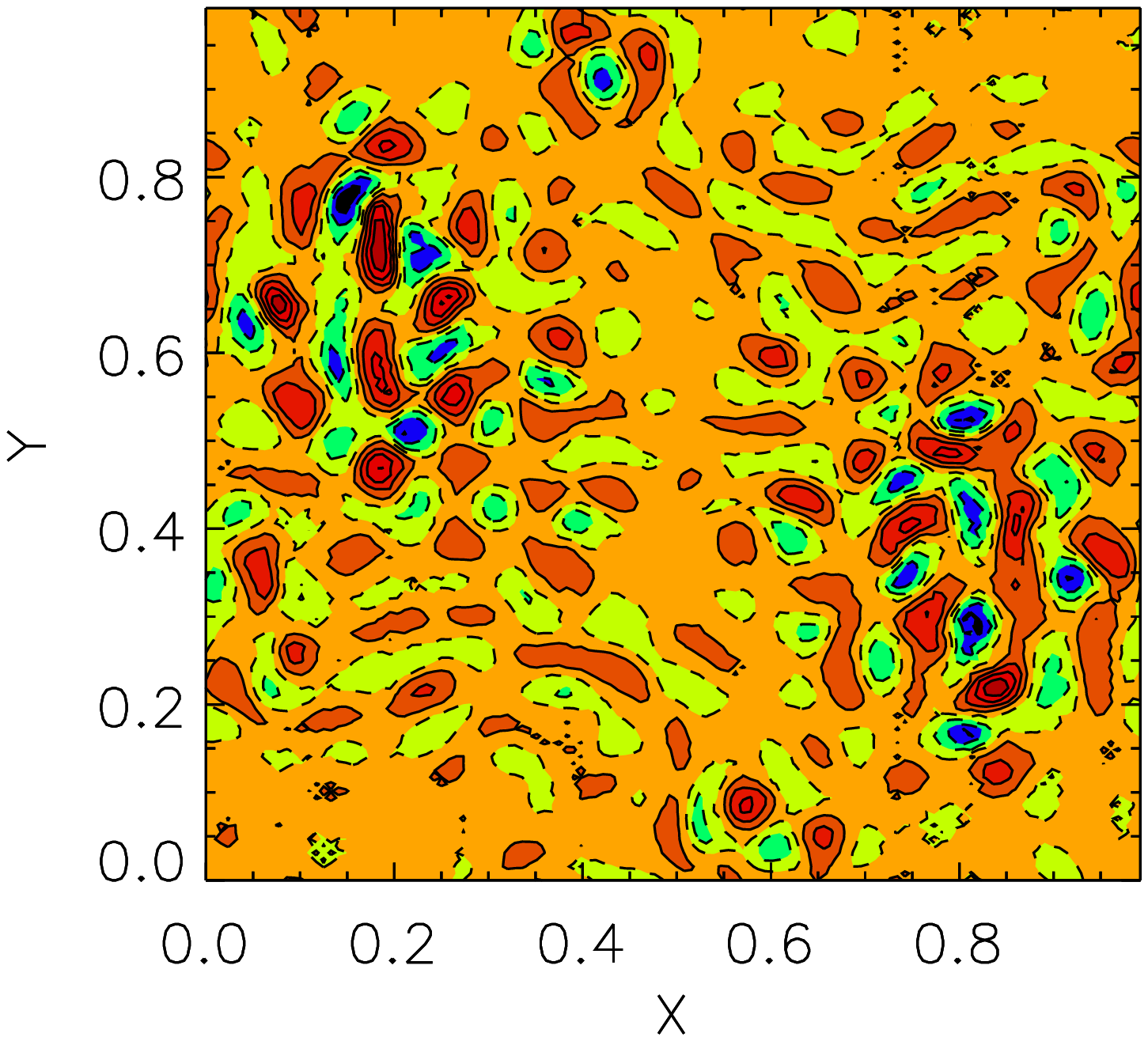}
\end{center}
\label{Fig3}
\end{minipage}\hfill
\vskip 1cm
\caption{Distribution of $V_z$-component of the velocity field with ranges $(-675,\, 701)$, $(-153,\, 157)$}
\end{figure*}

\begin{figure*}[t]
\vskip -4cm
\begin{minipage}[h!]{.45\linewidth}
\vspace*{2mm}
\begin{center}
\vskip -30cm
\includegraphics[width=40cm]{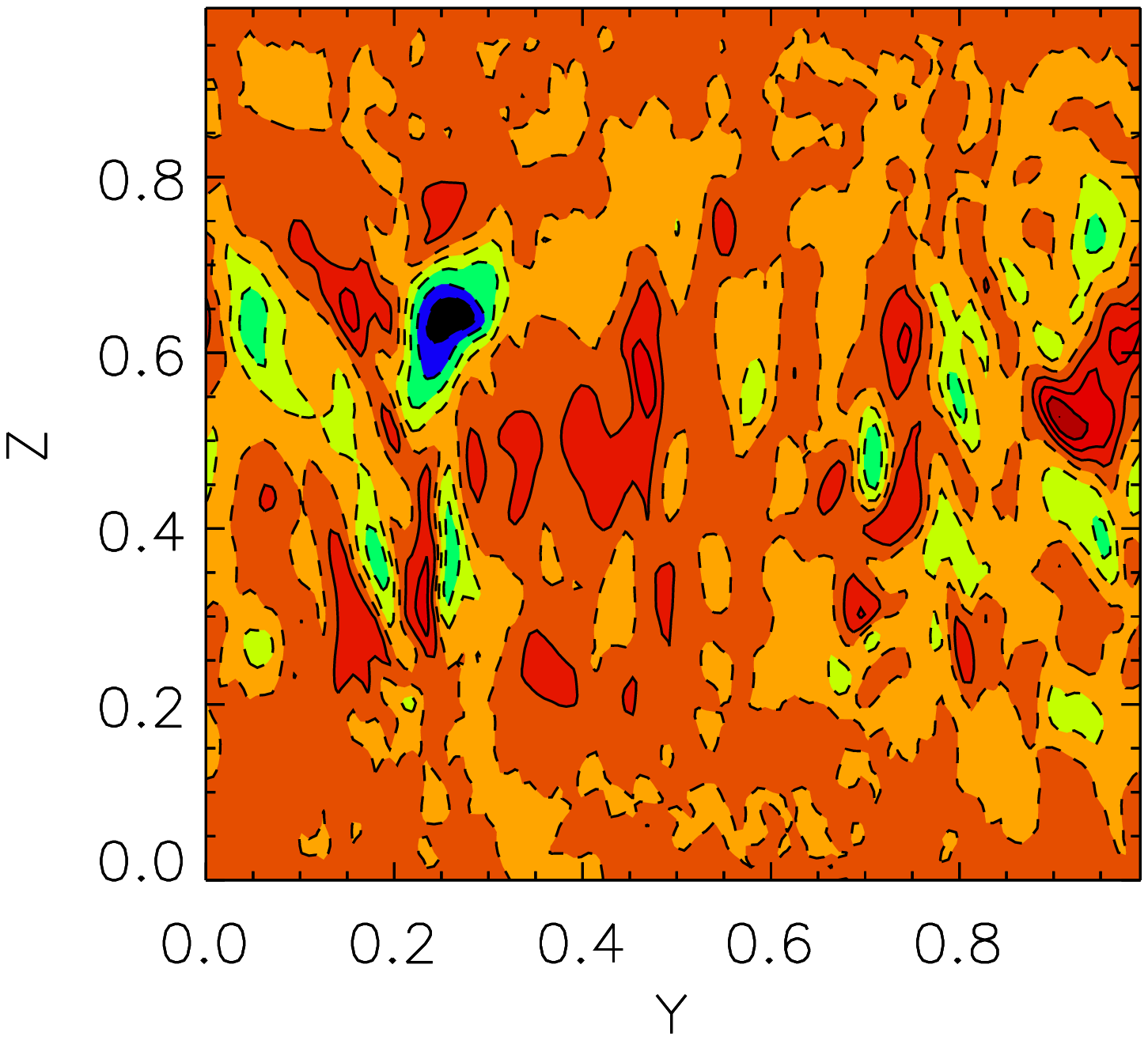}
\end{center}
\end{minipage}\hfill
\hskip -10cm
\begin{minipage}[h!]{.45\linewidth}
\vspace*{2mm}
\begin{center}
\vskip -30cm
\includegraphics[width=40cm]{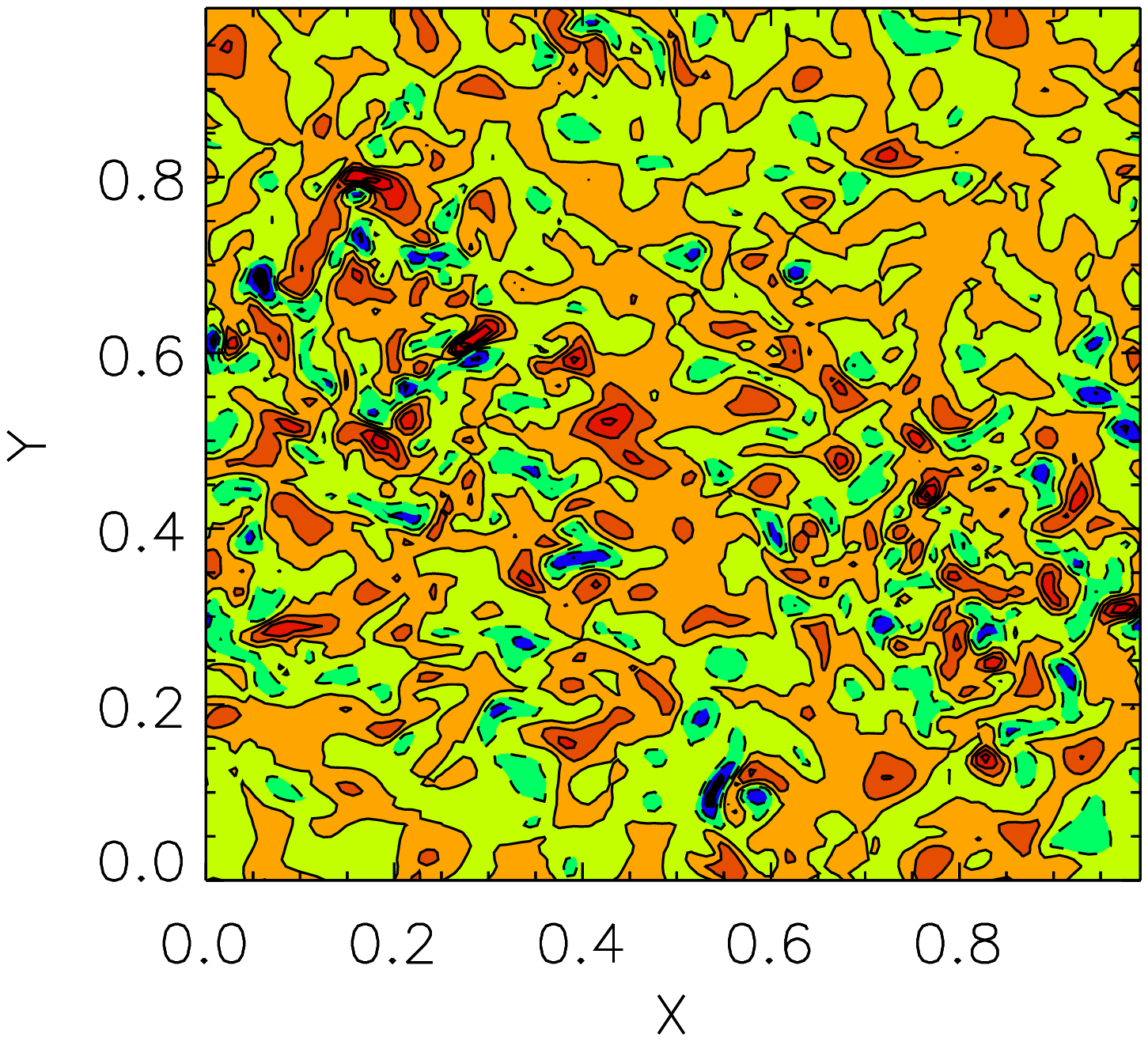}
\end{center}
\label{Fig4}
\end{minipage}\hfill
\vskip 1cm
\caption{Distribution of $B_z$-component of the velocity field with ranges $(-1.44,\, 1.14)$, $(-1.88,\, 2.37)$}
\end{figure*}

\begin{figure*}[t]
\vskip 0cm
\begin{minipage}[h!]{.45\linewidth}
\vspace*{2mm}
\begin{center}
\vskip -0cm
\hskip -1cm \includegraphics[width=8cm]{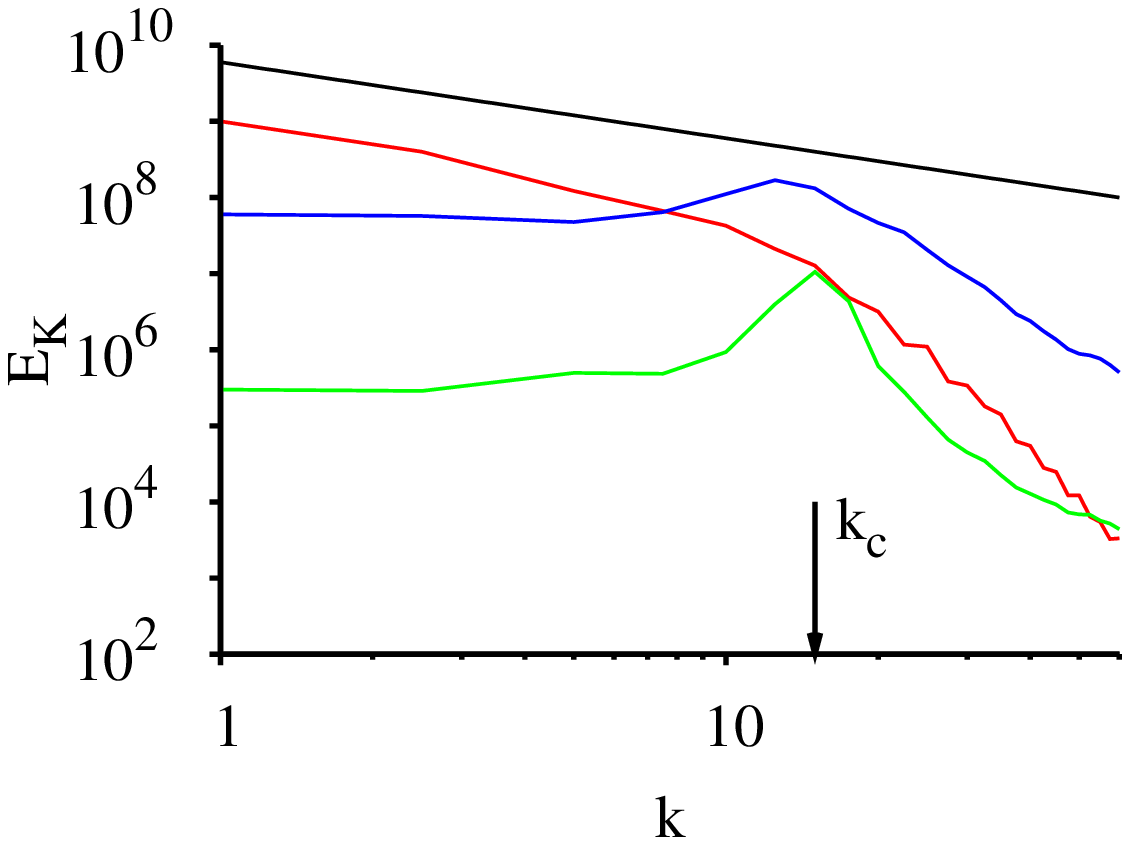}
\end{center}
\end{minipage}\hfill
\hskip -0cm
\begin{minipage}[h!]{.45\linewidth}
\vspace*{2mm}
\begin{center}
\vskip -0cm
\hskip -0cm\includegraphics[width=8cm]{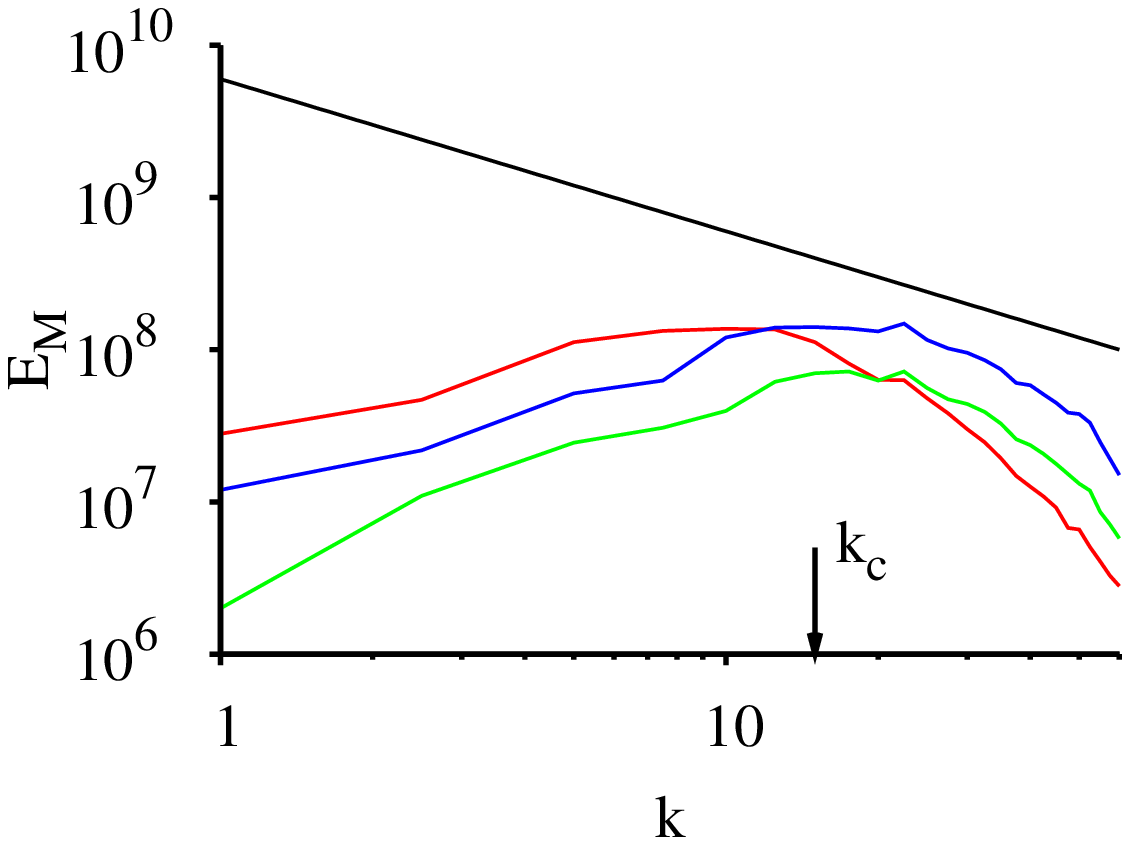}
\end{center}
\label{Fig5}
\end{minipage}\hfill
\vskip -0.5cm
\caption{On the left
 is spectra of the kinetic energy for  NR (red), R1 (green), R2 (blue).
    On the right is a spectra of the magnetic energy. The straight  line corresponds to the Kolmogorov's spectrum
  $\sim k^{-5/3}$.}
\end{figure*}

\begin{figure*}[t]
\vskip -0cm
\begin{minipage}[h!]{.45\linewidth}
\vspace*{2mm}
\begin{center}
\vskip -0cm
\hskip -1cm \includegraphics[width=8cm]{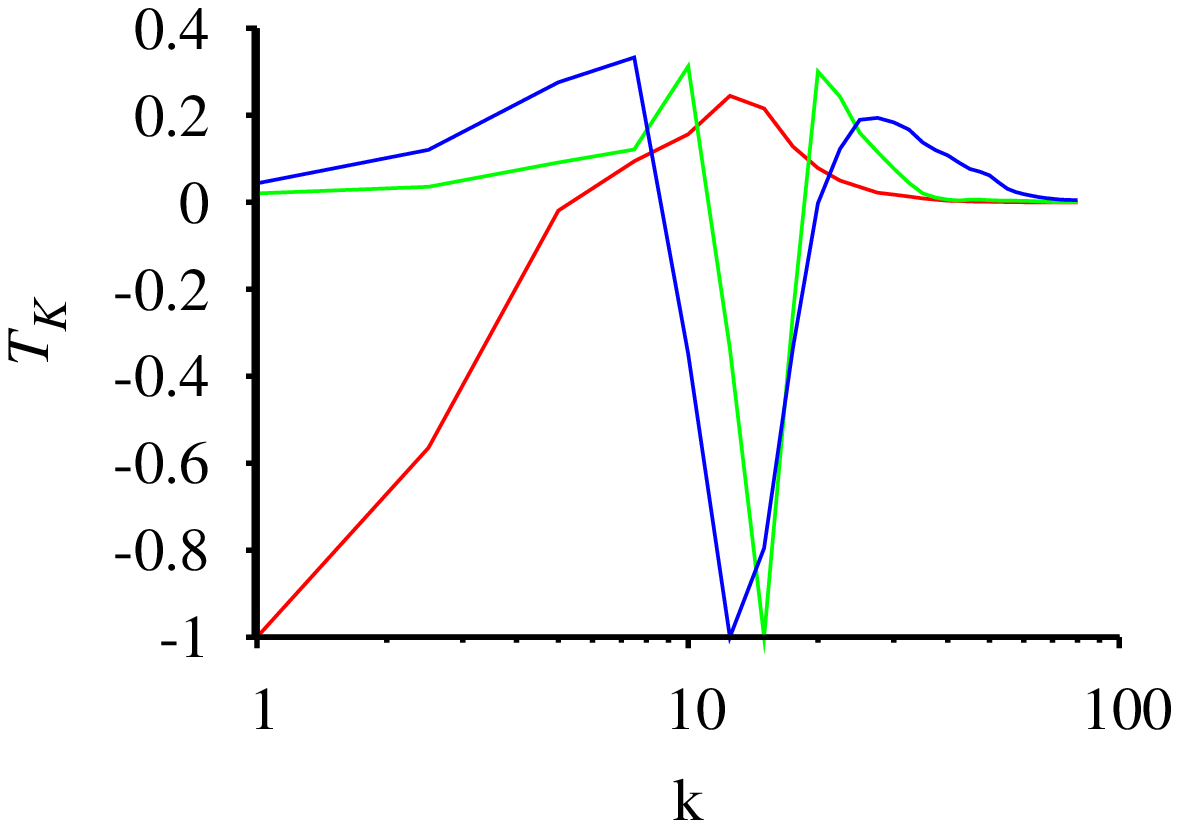}
\end{center}
\end{minipage}\hfill
\hskip -0cm
\begin{minipage}[h!]{.45\linewidth}
\vspace*{2mm}
\begin{center}
\vskip -0cm
\hskip -0cm\includegraphics[width=9cm]{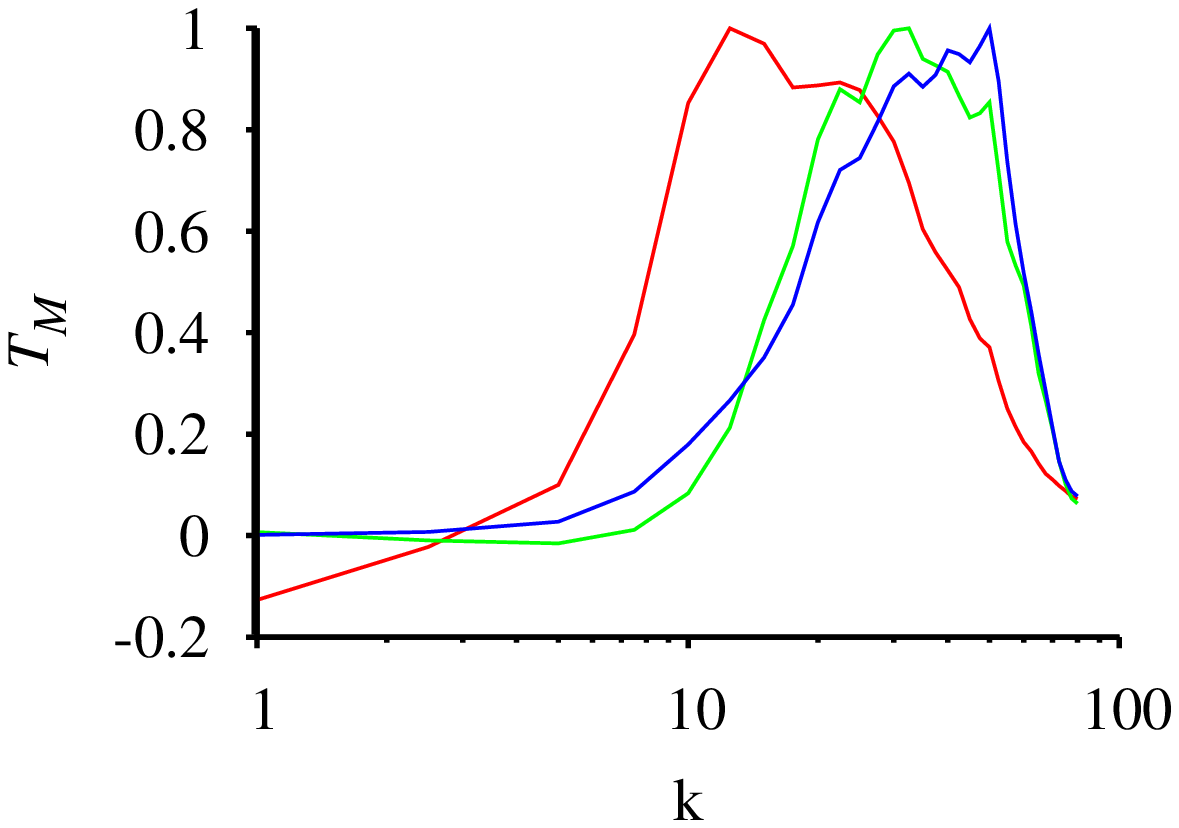}
\end{center}
\label{Fig6}
\end{minipage}\hfill
\vskip -0.5cm
\caption{The normalized fluxes of kinetic  ${ T}_K$
(on the left) and magnetic  ${ T}_M$  (on the right) energies in the wave space: 
NR (red), R1 (green), R2 (blue). 
%The values of the absolute extremums for the three regimes:
%  \\
% $-2.3\cdot 10^{12}$, $-8.3\cdot 10^{10}$, $-2.4\cdot 10^{13}$ (left).
% $ 2.0\cdot 10^{11}$, $2.3\cdot 10^{6}$,   $8.1\cdot 10^{7}$   (right).
}
\end{figure*}

\begin{figure*}[t]
\vskip -0cm
\begin{minipage}[h!]{.45\linewidth}
\vspace*{2mm}
\begin{center}
\vskip -0cm
\hskip -1cm \includegraphics[width=8cm]{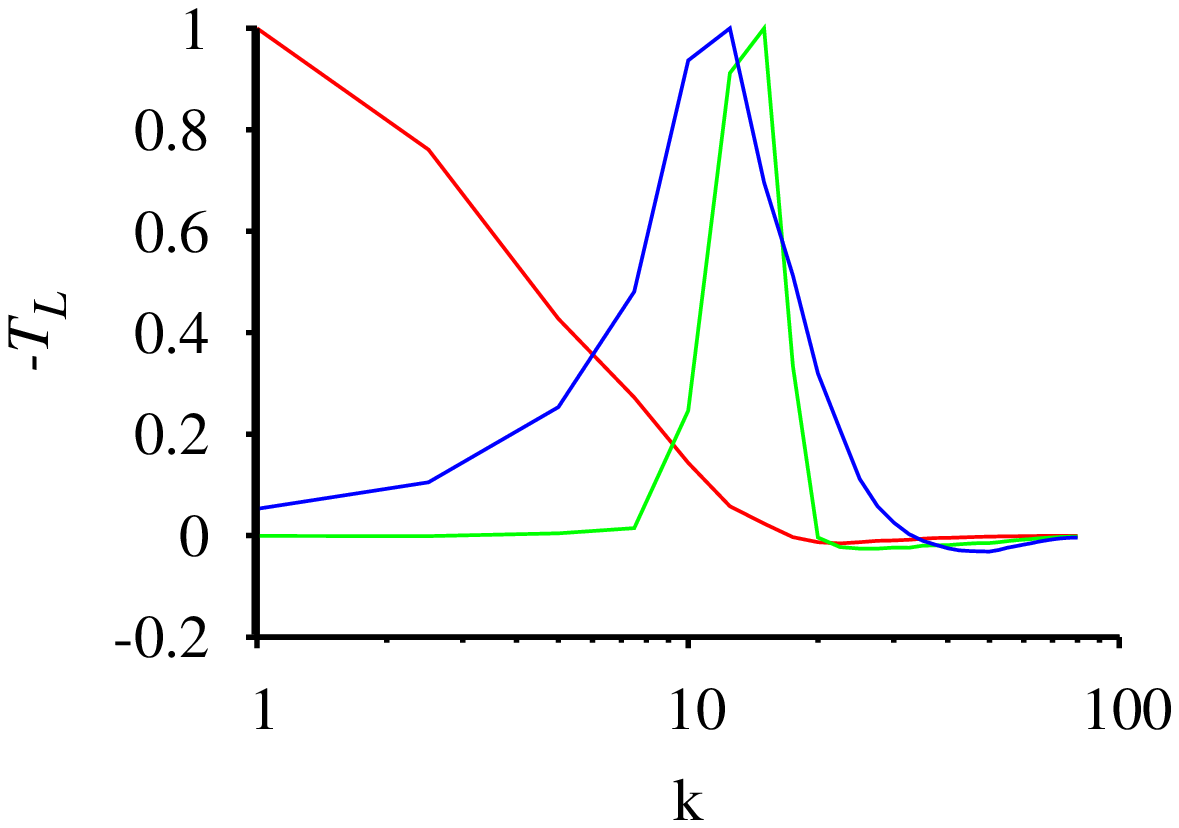}
\end{center}
\end{minipage}\hfill
\hskip -0cm
\begin{minipage}[h!]{.45\linewidth}
\vspace*{2mm}
\begin{center}
\vskip -0cm
\hskip -0cm\includegraphics[width=8cm]{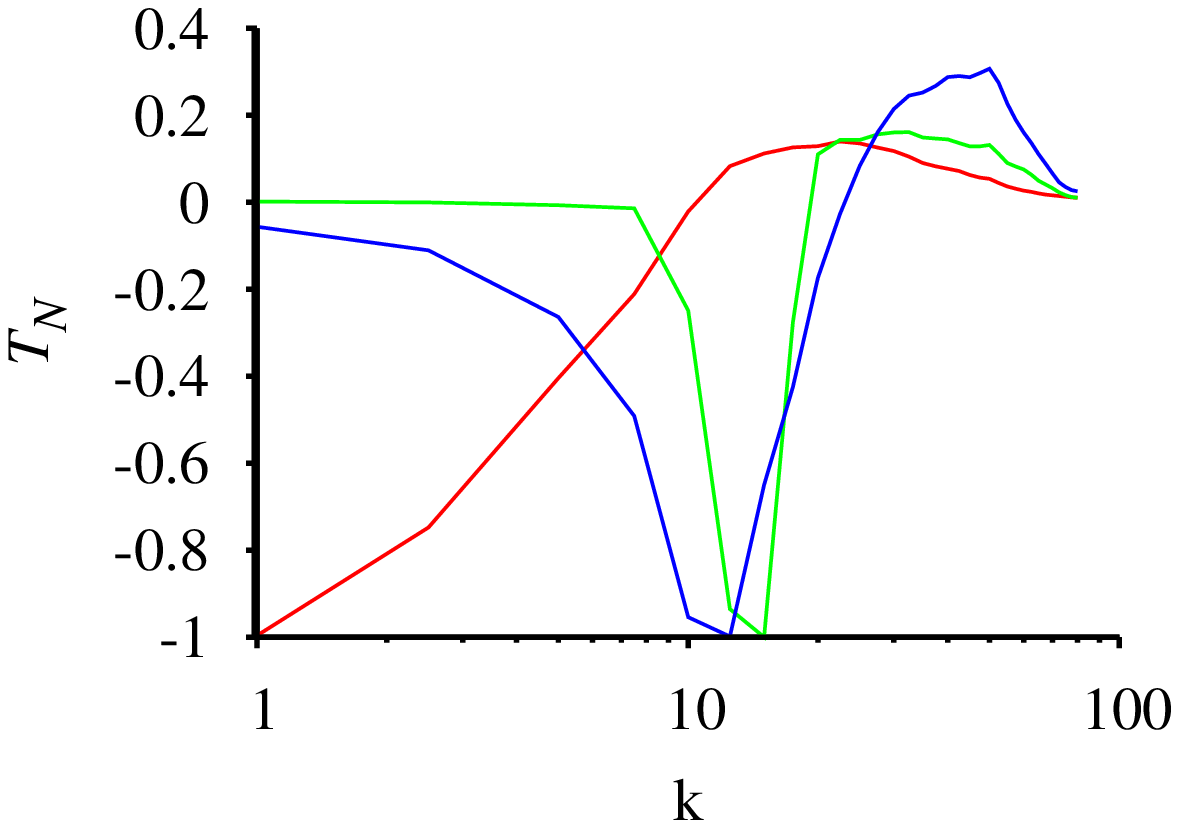}
\end{center}
\label{Fig7}
\end{minipage}\hfill
\vskip -0.5cm
\caption{
The normalized fluxes of the generation term   $-{ T}_L$
(on the left) and advective c  ${ T}_N$  (on the right) terms in the wave space: 
NR (red), R1 (green), R2 (blue). 
%The values of the absolute extremums for the three regimes:
%  \\
% $-2.3\cdot 10^{12}$, $-8.3\cdot 10^{10}$, $-2.4\cdot 10^{13}$ (left).
% $ 2.0\cdot 10^{11}$, $2.3\cdot 10^{6}$,   $8.1\cdot 10^{7}$   (right)
}
\end{figure*}

\section{Basic properties of the fields}
 Here we consider simulations without rotation similar to \cite{Meneguzzi} and with rotation for two regimes
   with different
amplitudes of the heat sources:
\begin{enumerate}
\item[NR:] Regime without rotation, $\Ra=6\cdot 10^6$, $\Pr=1$, $\E=1$, $\q=10$,  $\Re\sim 2.5\cdot 10^3$.

\item[R1:] Regime with rotation, $\Ra=1.3\cdot 10^3$, $\Pr=1$, $\E=2\cdot 10^{-6}$, $\q=10$, $\Re\sim 1.6\cdot 10^3$.

\item[R2:] Regime with rotation, $\Ra=2.1\cdot 10^3$, $\Pr=1$, $\E=2\cdot 10^{-6}$, $\q=10$, $\Re\sim 3\cdot 10^3$.

\end{enumerate}
\noindent
The first (NR) regime is close to the typical Kolmogorov convection, see in more details \cite{Meneguzzi}. Including
 of rotation (regime R1) Fig.~1-5 leads to transform of the isotropic  convective structures to the cyclonic state with
  horizontal scale $\sim \E^{1/3}$ ($k_c\sim \E^{-1/3}$) \cite{Ch}. Including of the magnetic field
  (the full dynamo regime
   with magnetic energy comparable with the kinetic energy on the order of magnitude)
   does not change structure of the convective patterns too much \cite{Jones}.
     In the same time spectra of the magnetic energy are
     quite different and have no well-pronounced  maximum  at $k_c$.

     Increase of Rayleigh number leads to decrease of the relative role of rotation and should decrease peak at the
      kinetic spectra energy what is accordance with spectra for regime R2, Fig.~5. In principle further increase
       of $\Ra$ should lead to the original Kolmogorov state, similar to NR with
        spectrum law $\sim k^{-5/3}$ Fig.~5. However we emphasize, that  information on the
        spectra is not enough to judge if the role of rotation is negligible or not and additional analysis is needed.
         The argument is the follows: rotation leads to degeneration of the third dimension (along z-axis) \cite{Batch}.
          On the other hand in isotropic two-dimensional systems spectraum of the kinetic energy also have $-5/3$-slope,
           however direction of the energy transfer in the system is inverse. In contrast to the three dimensional
            turbulence where energy transfers from the small wave number, where energy is injected, to larger dissipative
             wave number, in two dimensional turbulence energy transfers from the large wave numbers to the small
             ones\footnote{There is also a direct cascade of enstrophy.}. So far the quasi-geostrophic turbulence
             inherits properties of the both systems 2D and 3D, the inverse cascade \cite{Hossain, Constantin},
              we plan to consider behavior of
              energy fluxes in the wave space more carefully.

\section{Energy fluxes}
To analyze  energy transfer in the wave space we follow \cite{Frisch}. Decompose physical field
 $f$  in sum of low-frequency and high-frequency counterparts:
      $f({\bf r})=f^<({\bf r})+f^>({\bf r})$, where
      \begin{equation}\begin{array}{l}\dsize
f^<({\bf r})=\sum\limits_{|k|\le K} \widehat{f}_k\,e^{i{\bf k}{\bf r}}, \qquad
f^>({\bf r})=\sum\limits_{|k|>K} \widehat{f}_k\,e^{i{\bf k}{\bf r}}.
\end{array}\label{sys1}
\end{equation}
For any periodical  $f$ and  $g$ one has relation  \cite{Frisch}:
  \begin{equation}\begin{array}{c}\dsize
<{\partial f \over\partial x}>=0,\qquad <{\partial g \over\partial x}>=0, \\ \\  \dsize
<g{\partial f \over\partial x}>=
-<f{\partial g \over\partial x}>,\qquad
<f^>g^<>=0,
\end{array}\label{sys2}
\end{equation}
where
\begin{equation}\begin{array}{c}\dsize
<f({\bf r})>={\cal V}^{-1}\int\limits_{\cal V} f({\bf r})\, d{\bf r}^3
\end{array}\label{sys3}
\end{equation}
means averaging of
 $f$ over the volume  ${\cal V}$.
Multiplying the Navier-Stokes equation by ${\bf V}^<$ and induction equation by ${\bf B}^<$
 lead to equations of the integral fluxes of kinetic $E_K=V^2/2$ and
magnetic $E_M=B^2/2$ energies from $k\ge K$  to  $ k<K$:
\begin{equation}\begin{array}{c}\dsize
\Pi_K=<\left({\bf V}\times {\rm rot}{\bf V}\right)\cdot {\bf V}^<>,\qquad \\ \\ \dsize
\Pi_M=<{\rm rot}\left({\bf V}\times {\bf B}\right)\cdot {\bf B}^<>
\end{array}\label{sys4}
\end{equation} and
 for the flux of the  Lorentz work:
\begin{equation}\begin{array}{l}\dsize
\Pi_L=<\left({\rm rot}{\bf B}\times {\bf B}\right)\cdot {\bf V}^<>.
\end{array}\label{sys44}
\end{equation}
Introducing
      \begin{equation}\begin{array}{l}\dsize
     {   T}_K(k)= -{\partial \Pi_K \over\partial k},
\end{array}\label{sk}
\end{equation}
leads to obvious relation for $E_K$  in $k$-space:
\begin{equation}\begin{array}{l}\dsize
 {\partial E_K(k)\over\partial t}
=
{   T}(k)+F(k)+D(k),
\end{array}\label{s1}
\end{equation}
where $k=|{\bf k}|$,  ${   T}(k)$ is a flux of the energy from harmonics with
 different  $k$,
$F(k)$ is a work of the external forces and
$D(k)=-k^2E_K(k)$ is a dissipation.
The accurate form of  ${ T}$ is:
\begin{equation}\begin{array}{l}\dsize
{ T}_K=-{\partial \Pi_K \over\partial k}, \qquad \dsize
     \int\limits_{k=0}^{\infty} { T}_K(k)\, dk=0, \\  \dsize
{ T}_M={\partial \Pi_M \over\partial k},  \qquad
{ T}_L={\partial \Pi_L \over\partial k}.
\end{array}\label{sys5}
\end{equation}
Taking into account that
${\rm rot}\left({\bf V}\times {\bf B}\right)=-\left({\bf V}\cdot \nabla \right) {\bf B}+
\left({\bf B}\cdot \nabla \right) {\bf V}$ one has:
${ T}_M={ T}_N-{ T}_L$, where
\begin{equation}\begin{array}{l}\dsize

{ T}_N=-{\partial \Pi_N \over\partial k}, \qquad \dsize
     \int\limits_{k=0}^{\infty} { T}_N(k)\, dk=0, \\
\Pi_N=<
\left(\left(
{\bf V}\cdot \nabla \right) {\bf B}^<\right)\cdot {\bf B}^<>.
\end{array}\label{sys6}
\end{equation}
Fig.~6 presents fluxes of kinetic   ${ T}_K$  and magnetic
  ${ T}_M$ energies for the mentioned above regimes.
 Regime  NR  for  ${ T}_K$ demonstrates well-known behaviour for the direct Kolmogorov's cascade in 3D.
 For the large scales
   ${ T}_K<0$, these scales are donors, they provide energy to the system. On the other hand harmonics at
   the large $k$ absorb energy. The two-dimensional turbulence exhibits  mirror-symmetrical behaviour relative to
    the axis of absciss  \cite{Kr}. In this case the energy cascade is inverse.

Rotation changes behaviour of fluxes of kinetic energy essentially.  The leading order wave number is
   $k_c$. For  $k>k_c$ we also observe the direct cascade  of energy ${ T}_K>0$. The maximum of
  ${ T}_K$ is shifted relative to the maximum of the energy to the large $k$
   as larger as larger $\Re$.
    For  $k<k_c$ behaviour is more complex: for the
 small $k$ the inverse cascade of the kinetic energy takes place  ${ T}_K>0$.
 On the other hand for the larger region of $k$
 $(0\dots k_c)$ we still have the direct cascade ${ T}_K <0$. Increase of
  $\Re$  leads to narrowing of the region with inverse cascade and increase of inverse flux. One can
   suggest, that change of the sign of the flux
 ${ T}_K$ at  $k<k_c$ is connected with appearance of the non-local energy transfer:
   so that energy to the large-scales  ${\bf k_1}$
  comes from modes
$|{\bf k_2}|\sim |{\bf k_3}|  \gg |{\bf k_1}|$,
  ${\bf k_1}={\bf k_2}+{\bf k_3}$ \cite{Waleffe}. In absence of the magnetic field  maximum of
   ${ T}_K(k=1)$ appears. So, in case with rotation there are two cascades of kinetic energy (direct and inverse)
     tale place simultaneously.

Now we consider the magnetic part.
 In contrast to   ${ T}_K$, ${ T}_M$ includes not only advective term
  but the generative term as well. This leads to positiveness    of the integral  ${ T}_M$ over all $k$.
   Moreover   ${ T}_M$ is positive for any $k$. Position of maximum
    of    ${ T}_M$ is close to that ones in spectra of
 $E_M,\, { T}_K$.

It is evident, that for the planetary cores distance between maxima in fluxes
 ${ T}_M$ for  NR and R1, R2 can be quite large, however not so large as
  $k_c$. This statement concerned with condition on magnetic field generation which holds when
   the local magnetic Reynolds number ${\rm r_m}>1$ at the scale
 $1/k$: $\dsize {\rm r_m}= {v_k\over k \eta}>1$ and that for the planets  $\eta\gg
 \nu$. In the same moment fluxes at the small  $k$ are small, i.e. system is
 in the state of the statistical equilibrium: dissipation at the small scales is negligible.

Now we examine the origin of the magnetic energy at the scale $1/k$: does it concerned with the energy transfer from
 the other scales either it is a product of real generation at this scale?

Fig.~7 demonstrates fluxes of  $-{ T}_L$, concerned with magnetic field generation.
 The maximum of generation term without rotation is at the large scale, while for the rotating system it is at
  $\sim 1/k_c$. Interestingly, that for the rotating system there is a region $-{ T}_L<0$ for the large $k$, where
   magnetic field reinforce convection. For regime NR  $-{ T}_L$ drops quickly because of the kinetic energy decrease
     Fig.~5.
  As a result we have: for the rotating system magnetic field is produced by the cyclones, while for the non-rotating system
   the large-scale dynamo operates.

Now we estimate the role of the advective term ${ T}_N$ separately. For the non-rotating
    ${ T}_N$ and
    ${ T}_K$ are similar: the direct cascade takes place.
  For rotating system region
 $k\sim k_c$ is a source of energy.
 In contrast to
 ${ T}_K$, ${ T}_N$ has not positive regions at the small  $k$, i.e. the inverse cascade
  of the magnetic energy, concerned with the advective term in this region is absent.
 We draw attention to the amplitudes of the fluxes
 ${ T}_M$, $-{ T}_L$, ${ T}_N$:  for all three cases 
hold  $\dsize {{| T}_M|\over |{ T}_L|}\sim  10^{-1}$, i.e. two fluxes of the magnetic energy in the wave space
  with opposite directions exist. The first flux concerned with
 the traditional energy transfer of the energy over the spectrum (advective term) and the flux of the Lorentz work.
 The regions of the maximal magnetic field generation coincide with the regions of the most effective
 magnetic energy transfer
 ${ T}_N$ (from small $k$ to the large  $k$). Such a balance leads to equipartition state, when dissipation
  takes place on the large $k$.

Note, that the full magnetic flux
Fig.~6  ${ T}_M$ is localized at  $k\gg 1$.
 For the non-rotating system it is because the mean helicity and  $\alpha$-effect are zero \cite{Zeldovich}.
 Thus the inverse cascade of the magnetic energy to the small $k$ is absent.

For the rotating system here is a balance of the energy injection due to the Lorentz force and its wash-out because of
 advection. The latest effect reduces $\alpha$-effect.

\conclusions
%% \conclusions[modified heading if necessary]
The magnetic fields of the planets are the main sources of information on the
 processes in the liquid cores at the times
 $10^2-10^3$yy and more.
  If the poloidal part of the magnetic field is observable at the planet surface, then the
   largest component of field (toroidal) as well as the kinetic energy distrubution
   over the scales is absolutely invisible for the observer out of the core.
 Moreover, due to finite conductivity of the mantle
   even the poloidal part of the magnetic field is cutted off at $k\ll k_c$.
 In other words the observable part of the spectra at the planets surface is only
 a small part (not even the largest) of the whole spectra of the field.
    That is
   why importance of the numerical simulation is difficult to overestimate. Here we showed
   that in considered  quasigeostrophic state
 the both cascades (direct and inverse) exsist simulataneously. That is a challenge for a turbulent models in the geodynamo. 
The other interesting point is 
 a balance of the magnetic energy flux due two mechanisms: advection and generation. This balance is reason why
  the exponential growth of the magnetic energy is stopped at the end of the kinematic dynamo regime.

%\appendix
%\section{\\ \\ \hspace*{-7mm} HEADING}    %% Appendix A

%\subsection{asdfsfds}                               %% Appendix A1, A2, etc.

%\begin{acknowledgements}
%TEXT
%\end{acknowledgements}

\end{document}